\documentstyle[prl,aps,multicol,epsf]{revtex}
\begin{document}

\draft
\tighten

\title{Current ideas on the origin of stripes.}

\author{J. Zaanen}
\address{Institute Lorentz for Theoretical Physics, Leiden University\\
P.O.B. 9506, 2300 RA Leiden, The Netherlands}
\date{\today ; E-mail: jan@lorentz.leidenuniv.nl}
\maketitle

\begin{abstract}
It will be attempted to present a coherent view on the current ideas 
regarding the origin of the stripe instability. Special emphasis will be
put on the problem of how to combine the microscopic pictures, leaning on
spin-charge topological aspects, with the notion of frustrated phase
separation.
\end{abstract}

\begin{multicols}{2}
\narrowtext
I have noticed that newcomers to the field of stripes\cite{tranquada}
seem to percieve an atmosphere of theoretical controversy surrounding 
the issue of stripe microscopy.  It is actually so that there is a sense of
growing consensus on what is understood -- self-evidently, much remains still 
in the dark. The confusion finds its origin in the history of the subject:
stripes were several times (re)discovered theoretically over the past ten years
from quite different physical perspectives. It took a little time to
realize that these different pieces of physics in fact all matter.

I will present here a crude sketch of the main ideas, ordered according
to my personal view on this subject.
Stripes were discovered for the first time by Gunnarsson and
me\cite{zagu} 
as a generic classical instability of doped Mott-Hubbard insulators, with
classical in the sense of `integrating out fermions around the classical
saddlepoint' -- lingo for Hartree-Fock. Although still of relevance,
subsequent developments made clear that naive Hartree-Fock does miss
some essential pieces of cuprate physics. I refer in the first
place to the highly non-trivial role played by quantum 
fluctuations (section I): 
the realization by Prelovsek and coworkers\cite{prelov}
that the stripe instability emerges entirely from the soup of quantum
fluctuations, 
linking the phenomenon to spin-charge separation; the very recent discovery
by White and Scalapino\cite{whisc}
that stripes can be made out of pairs of holes, instead of single holes,
suggesting interesting relationships with superconductivity.

A theorem by Laughlin states that a complicated theory which is right
looses on the short term from a simple theory which is 
wrong\cite{laughl}. I percieve the statement that 
stripes exist {\em because} there are short range
attractive interactions and long range repulsive interactions as a theory
of the second kind. It is even so that it can be
directly seen from the experiments that such a statement does not make sense.
This is different from claiming that longer range interactions (neglected
in Hubbard models) do not exist. The frustrated phase 
separation mechanism\cite{emkiv}
hits full force at $x > 1/8$ where the stripes likely become internally
charge compressible (section II).

\section{Stripes as holons in two dimensions.}

A first set of ideas emerged from microscopic calculations. They have
in common that stripes can be looked at as 2D generalizations of the 
topological excitations known from 1D physics. One can
excercise the  notion that the holons from one dimensions do survive in 2D 
as long as they bind into stripes\cite{nayak}. However, this should be
handled with care: as the remainder will make clear, stripes have to be an
irreducible two dimensional phenomenon which cannot be simply thought of
as $N$ times one dimensional physics. I actually like to dream that stripes
are about nature teaching us how to properly generalize the mathematical
theory of one dimensional physics to higher dimensions. This theory is
yet to be discovered, and all we posses at the moment are some vague 
contours of the real thing. 
 
The lesson to be drawn from the early mean-field calculations is
that the problem of the doped Mott-Hubbard insulator is on the semi-classical
level a close relative 
of the doped Peierls insulator. In the latter, the doped holes bind to
topological defects (domain walls) of the density wave, forming 
the so-called Su-Schrieffer-Heeger solitons\cite{ssh}. 
Although the doped Mott-Hubbard
insulator is different (the antiferromagnet is itself made out
of electrons and the order parameter is a vector), 
this soliton mechanism was shown to survive in this 
context\cite{zagu,morewalls,inui},
with the longitudinal N\'eel order parameter component taking the role
of the phonon field of the SSH problem. As was substantiated by 
Zaanen and Ole\'s\cite{zaol},
the stability of the charged domain walls is best understood by first
considering the hole motions perpendicular to the walls. This is like a
one dimensional problem, where the order parameter defect pulls out 
a `mid-gap' state from the Hubbard bands which is occupied by the
carrier. This corresponds with a hole bound to an Ising-like domain wall,
in the sense that the staggered order parameter changes sign upon passing 
the localized charge, regardless its overall orientation (`classical holon').
In two dimensions,
these holons can only survive when they are `put on a row' to satisfy 
the topological requirements of a 2D N\'eel order parameter. Remarkably,
the energetics of at least the filled wall (one hole/domain wall unit cell)
is one dimensional-like in the sense that most of the energy is gained 
by the motions of the holes perpendicular to the wall. For instance, these
Hartree-Fock walls are characterized by an extremely soft transversal
dynamics: as long as the holons form a connected trajectory, the energies
associated with shape deformations of the walls are barely detectable.
In a sense, this is like `$N$ times one dimensional physics'. However,
it is a specialty of this particular type of stripe phase. The stripes
of relevance to cuprates are half-filled (one hole per domain wall unit
cells), and as will be explained further in the next section, in order
to acquire special stability a gap has to develop in the mid-gap band
itself\cite{zaol,nayak1}. Since the physics of a system of particles on a 
line is not reducible to that of particles living on disconnected points,
these half-filled stripes are truely 2D objects.

\begin{figure}[h]
\hspace{.2 \hsize}
\epsfxsize=.5 \hsize
\epsffile{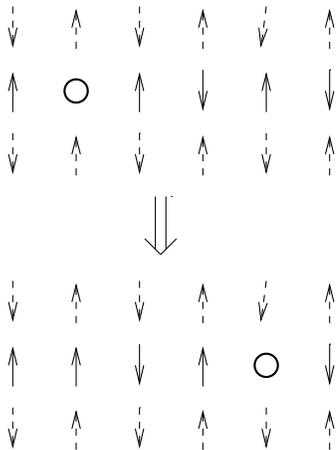}
\caption{In one dimension, charge-spin separation finds its origin in
simple kinematics: a hole injected at the origin has transformed after
a couple of hops into a domain wall-like spin defect (spinon) and a hole
surrounded by antiparallel spins (holon). In two dimensions (dashed arrows),
the spinon and holon are confined by the magnetic string potential: the
spins are flipped along the trajectory connecting the spinon and the holon,
and the number of parallel spins grows proportional with the distance 
between the two.}
\end{figure}

Despite their meat-and-potato image, classical saddlepoints are very serious
objects. Since the static stripe phase is a classical vacuum (a
N\'eel state and a charge density wave), there has to be a theory which
is controlled by this vacuum, and this theory has to have the structure
of Hartree-Fock: the quasiparticles can be integrated out by letting them
scatter against the order parameter potential, and this zero-th order
state can be adiabatically continued to the true vacuum. The caviat is
of course that this theorem is only valid with regard to the appropriately
renormalized Hamiltonian. There is no a-priori reason to believe that
simple Hubbard models can claim this status. Nevertheless, 
it seems that Hubbard model Hartree-Fock gives a correct
and even quantitatively meaningful description
of the ground state of doped nickelates, where the stripes were first observed\cite{tranquadani,zali}.

That it also can go wrong 
is vividly illustrated by the more recent discovery of stripes
in the $t-J$ model. It is easy to check that in terms of the bare spins
and bare holes, there is no stripe instability in the classical
($S \rightarrow \infty$) limit of the $t-J$ model. The $t-J$ model
has to do with the large $U$ limit of the Hubbard model, and it was
early on established\cite{inui} that stripes disappear on the mean-field level
when $U$ becomes larger than twice the bandwidth. {\em The stripes
of the $t-J$ model are a genuine quantum 
order-out-of-disorder phenomenon.} Without quantum fluctuations ($S \rightarrow
\infty$) stripes do not exist, and one has to go far beyond the Gaussian
level (linear spin waves) to recover the stripe instability. In fact, 
Prelovsek and coworkers, who were the first to identify a tendency for
stripe formation in the $t-J$ model\cite{prelov}, came up with a mechanism
linking it to the holons of one dimensional physics.

In one dimensions, spin-charge separation is 
a trivial kinematical effect and it can be illustrated by a simple strong
coupling cartoon\cite{schulz}. A bare hole can freely hop in a $S=1/2$
spin system, and after a couple of hops 
the hole has `decayed' into a Ising-like domain `wall'
(better viewed as a Jordan-Wigner fermion; the spinon) while the hole is surrounded by anti-parallel spins: the hole is `bound' to a domain wall
(Fig. 1). Although this holon looks similar to the `classical
holon' of the Hartree-Fock solutions, it exists because of the 
delocalization of charge which is not present in the semiclassical theory.
This kinematical effect
is unavoidable in any clean one dimensional system and spin-charge separation
is physical law. In two dimensions, the spinon and holon
are connected by a `magnetic string' of flipped spins (Fig. 1):
the energy grows linearly with the spinon-holon separation because of the 
ferromagnetic bonds to the spins neighboring the string. At least in the 
well understood one hole case, the short time dynamics is spinon-holon 
like, but at low energies these confine to form a hole --
the well known Landauesque quasiparticle\cite{dagotto}.

Prelovsek and
coworkers\cite{prelov} recognized that in the many hole problem there 
is another possibility: holons do survive but they condense in connected
trajectories corresponding with fluctuating stripes. The underlying
mechanism adds a completely new meaning to the expression `electron
correlations'. Start out with
a stripe, viewed as a string of holons (Fig. 2). 
What happens when an individual holon
hops to a neighboring lattice site? The spin moves backward, but since
the stripe is an antiphase boundary, this spin ends up in a `right' spin 
domain, not causing any ferromagnetic bond. 
A neighboring holon can now move even more easily, and the net result 
is that two kinks propagate away freely, connected by a piece of 
stripe which is displaced by a lattice constant (Fig. 2)\cite{viertio}. 
Because all
holons can hop, these kinks will tend to proliferate, thereby delocalizing
the stripe as a whole. Hence, although the price is paid of a reduced
hopping configuration space, individual charges can hop as if there is no
antiferromagnetic background hindering their motions, by coordinating these
hoppings with those of {\em all} other particles. Prelovsek {\em et al.}
 substantiated this qualitative idea with extensive quantitative
calculations\cite{prelov}. 
Intrigued by these observations, we spend in Leiden much time on studying
further abstracted `lattice string' models. Given that kinks proliferate
as just described, how does the string as a whole fluctuate? This turned
out to be a rather amusing affair, with links to the hidden
order of Haldane spin chains, surface statistical physics, etcetera. The
tentative conclusion is a `don't worry theorem': when these strings
delocalize, their long wavelength dynamics is always of the simplest
possible kind, namely the one governed by free field theory\cite{eskes}.

\begin{figure}[h]
\hspace{.05 \hsize}
\epsfxsize=.8 \hsize
\epsffile{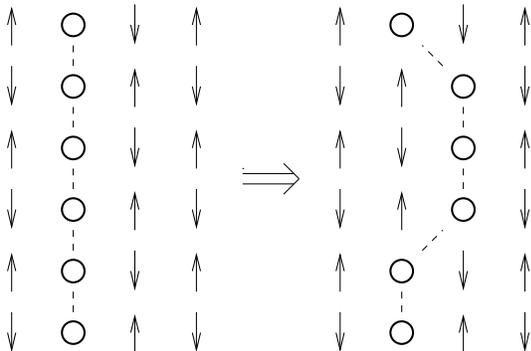}
\caption{The order-out-of-disorder mechanism of Prelovsek and Zotos.
The magnetic string potential can be avoided in 2D by forming closed 
trajectories of holons: when an individual hole hops, the spin moves
backwards into its right spin domain. The price is paid of a reduced
hopping phase space, but much of the kinetic energy is recovered by
the delocalization of the `string' as a whole.} 
\end{figure}

The exact diagonalization studies by White and Scalapino\cite{whisc} show 
that the above story is still incomplete. Although the dust has not
settled yet, their message appears to be that the stripes in $S=1/2$ systems 
are made out of {\em pairs} of charges instead of single 
charges\cite{rice}. This is
appealing, if not only because it has to be explained how  superconductivity
enters the picture\cite{dhlee}. 
It is apparently so that in the considerations by
Prelovsek {\em et al} the local quantum spin fluctuation is underestimated.
These tend to stabilize local singlet spin pairs, especially so when holes are
in their neighborhood.  The simulations show that there is a 
tendency to form a bound state consisting of two holes and two spins. 
The holes tend to sit on
opposite sides of the elementary plaquette, while the two spins on
the other corners form a strong singlet bond along the diagonal of
the plaquette. Leaving it to Scalapino {\em et al.} to explain this pairing mechanism in more detail, it is easy to see that they cause
anti-phase N\'eel correlations in the spin background.
The singlets are formed along the {\em diagonal} 
of the plaquette, forcing anti-parallel spin correlation on the {\em same}
sublattice. If the surrounding spin system is near to its classical limit,
the singlet-triplet logic of e.g. the bilayer Heisenberg model\cite{coen}
 applies and
it follows that these pairs act like local antiphase magnetic boundaries
(`bosonic holons').

\section{Relevance of long range interactions.}

As already announced in the introduction, there is a second set of ideas
which do not fit in directly in the smooth flow of ideas alluded to
in the previous section: frustrated phase separation\cite{emkiv}. 
There is little
to be explained, since the beauty of the argument is in its simplicity: a
classical system of particles on a lattice, characterized by short range
attractions and long range repulsions, minimizes its energy by forming
linear domains of enhanced and reduced density.
In the context of stripes, it is sometimes argued that this is the big number
physics and in the space left behind after all this has
happened, subtleties can occur of the kind described in the previous section.
The problem with this kind of argument is that it relies on a percieved
detailed knowledge of the short wavelength physics, and it is a long time
experience (at least mine) that this regime is littered with messy 
intricacies\cite{notesr}. Instead, the tractable question 
is as to what extent long range interactions are consequential for the
long wavelength physics. For instance, under the rule of frustrated
phase separation, the Rome group\cite{castell} might well be on the right
track developing the relevant field theory. Longitudinal fluctuations
are expected to dominate: holes fluctuate from hole rich to hole poor regions
thereby decreasing the amplitude of the stripe order parameter. Alternatively,
when the `holon glue' dominates, the stripes disorder by
transversal (shape-changing) fluctuations and the string liquid 
ideas\cite{nayak1,eskes,zahovs}
become more natural. The truth is of course somewhere in the middle, and it
is a matter of high urgency to find out where this somewhere is.

Regardless the importance of long range interactions, it is not always realized
that the `holon-glue' of the previous section is also about big
numbers. A key consequence of section I is that stripes
are also anti-phase boundaries in the spin system -- a fact not addressed
at all by frustrated phase separation. Obviously, the strength of the exchange
interaction $J'$ connecting the spins on opposite sides of the domain wall
is intimately related to the tendency of the hole (or pair) to surround
itself with antiparallel spins
and this exchange interaction is actually quite large. The spin
waves have been measured in a nickelate stripe phase\cite{nispwa}, 
indicating that
$J'$ is smaller by no more than a factor of two as compared to the exchange
at half filling. In addition, the dynamical fluctuations in superconducting
cuprates remain incommensurate up to 
rather large temperature- and frequencies\cite{aeppli},
and this is only possible when $J'$ is at least as large. The implication
is that a lower bound to the strength of the `holon-glue' interaction is
actually known and this quantity is of order of the exchange interaction
at half-filling -- a big number.

As emphasized by Emery and Kivelson\cite{emkiv1}, 
the behavior of the stripe period as
function of doping gives away an important clue. It was already emphasized
that, since static stripes do form a classical vacuum, a theory with the
{\em structure} of Hartree-Fock should exist. It is  possible
to make deductions based merely on the structure of the theory: if
the effective Hamiltonian is characterized by only short range interactions,
special stability is only obtained if the order parameter potential is
such that it causes a gap in the electronic spectrum in at least part
of the Brillioun zone. Applied to stripes, the mere magnitude of the
spins in the magnetic domains ($> 0.3 \mu_B$ at $x=1/8$) implies that
a rather large gap is associated with the electrons moving in these domains.
The quasi 1D electronic subsystem living on the wall is more delicate.
If the instability lives in the diagonal (spin density and/or charge density)
channel, gaps can only occur when {\em the charge density is
commensurate with the underlying lattice.}  
For instance, starting from Hubbard models one finds a preferred density
of one hole per domain wall unit cell (filling fraction $\nu =1$) corresponding
with a gap between the completely empty mid-gap band and the lower Hubbard band.
Alternatively, special stability at $\nu = 1/2$ (one hole per two unit 
cells) implies the presence of either a   
$2k_F$\cite{zaol} or $4k_F$\cite{nayak1} density wave
instability on the stripes.

If on-stripe charge-commensuration dominates, every hole adds a fixed
length of domain wall and it follows directly that the domain wall spacing $d$
is inversely proportional to the number of holes $x$, while the 
incommensurability $\varepsilon \sim 1/ d \sim x$. In the nickelates
this rule is very closely obeyed at low temperatures\cite{nihit} 
and since $\nu =1$
Hubbard model Hartree-Fock appears as a sensible theory\cite{zali}.
The situation in the cuprates is more interesting\cite{tranquada}.
For doping levels $x < 1/8$, one finds again $\varepsilon \sim x$ 
with $\nu = 1/2$, indicating on-stripe charge commensurability.
However, for $x > 1/8$ the stripe period remains roughly constant,
indicating that the stripes become internally charge compressible.

The significance of the frustrated phase separation argument is that
it shows that long range interactions can change the basic rules of
the game. In the presence of these interactions, charge density waves 
can acquire special stability, regardless the response of the states 
at the Fermi level. Charge density waves can exist in
purely classical systems (e.g.,  the Wigner crystal).
Consider the interesting situation of competing
short range attractive- and long range repulsive interactions where, 
as a subtle compromise, the charges pile up 
in linear stripe-like domains. What happens when this system
is mildly quantized? Since the kinetic energy favors a
homogeneous state, the charges spread out thereby decreasing the magnitude 
of the charge modulation and likely also changing the ordering wavevector. However, it will do so in some smooth way, unrelated to the wavevectors spanning
the emerging Fermi-surface. The system becomes a metal with a charge
density modulation which is to zero-th order driven by the interactions alone.

The microscopic picture of the previous section 
and the principle exposed in the previous paragraph refer to completely
different aspects of the physics, and there is nothing forbidding that
they are both active at the same time. Consider the classical
limit but take instead of featureless local attractions a minimal way
of incorporating the holon idea,
\begin{eqnarray}
\label{korneel}
H \; & = & \; \sum_{\vec{i}\vec{\delta}} n_{\vec{i}} n_{\vec{i}+\vec{\delta}} 
\vec{M}_{\vec{i}} \cdot \vec{M}_{\vec{i}+\vec{\delta}} + \nonumber \\
 &  &  J_h\sum_{\vec{i}\vec{\delta}} ( 1 - n_{\vec{i}} )
n_{\vec{i}-\vec{\delta}} n_{\vec{i}+\vec{\delta}} 
\vec{M}_{\vec{i}-\vec{\delta}} \cdot \vec{M}_{\vec{i}+\vec{\delta}}.
\end{eqnarray}
$1 - n_{\vec{i}}$ is the number operator of the charged particle (the hole,
or the pair of White and Scalapino) on the $\vec{i}$ site of
the lattice while $\vec{M}$ denotes the direction
of the staggered order parameter.
This describes a classical Heisenberg magnet doped by `classical holons': 
for $J_h > 0$, individual charges want to be
coordinated with an antiparallel configuration of the staggered magnetization.
For the present argument it suffices to know that
the ground state for $J_h \geq 1$ is a charge commensurate but transversally
disordered stripe state, while  for smaller $J_h$ phase separation occurs
(Eq. (\ref{korneel}) defines a surprisingly complicated 
statistical physics\cite{carmine}). 
If long range charge-charge interaction is added, stripe long range order
is stabilized and for $J_h =0$ these stripes are precisely of the frustrated
phase separation kind.  The `holon interaction'
disfavors broad stripes and when $J_h$ becomes larger the narrow `holons on
a row' stripes recover. However, these stripes are now `doped' in the sense
that additional holons are incorporated in the form of transversal kink
defects carrying a net charge\cite{carmine}. 
The mechanism is straightforwardly understood:
in the frustrated phase separation limit ($J_h = 0$) the stripe period is set 
by a different mechanism (competition short- and long range interactions) than
in the $J_h \rightarrow \infty$ limit (charge commensuration) and doped
stripes appear as a compromise in the intermediate regime.

\section{An open ended story.}

There are reasons to believe that the understanding of even the `easy' static
stripes in the cuprates is still highly incomplete. Let me list some of the 
most obvious problems: (i) Why is there are a kink in $\varepsilon$ versus $x$
curve at $1/8$\cite{tranquada}? 
It is not easy to see why this should happen, given the wisdom
of the previous section: there is no obvious reason why the system should
switch {\em suddenly} from on-stripe charge commensuration ($ x < 1/8$) to
stripe-to-stripe commensuration ($x > 1/8$).   
(ii) Why is it that the resistivity increases only slowly  (like a
logarithm) below
the charge-ordering temperature? Stronger, why is the low temperature 
resistivity {\em smallest} at $x=1/8$, while it increases for 
both higher- and lower dopings\cite{buechner}? $x=1/8$ is the point of maximal 
commensuration and this is not at all reflected in transport properties. 
$\varepsilon \sim x$ should relate to the presence of some gap in the
electronic spectrum  and one would expect a strong asymmetry in 
the transport properties around $x=1/8$. This asymmetry is absent. 
(iii) How to incorporate properly superconductivity? It seems that
the superconducting and stripe phases are connected by a (near) second order
phase boundary, and it might even be that a coexistence 
(`antiferromagnetic supersolid') phase exists\cite{tranquada}. 
On the most general level,
the dynamical stripe correlations showing up in the superconductors
should be understood as reflecting this second order behavior: on
`short' (in fact, relatively large) length scales the system still remembers
that it could become a stripe phase. Conversely, it should be the case that 
the stripe phase is also a superconductor, which failed at the very last moment.
This raises some problems of principle. Focussing on the charge sector,
the stripe phase is best called a complex solid, and the quantum liquid 
crystal ideas of Kivelson {\em et al.}\cite{qulicr}
 effectively illustrate what this
can mean for the phase dynamics. The stripe phase is also an antiferromagnet.      Although it is unclear to me why one should worry about a $0.6 \mu_B$
antiferromagnet at $x=0$, knowing about the lingering $0.3 \mu_B$
antiferromagnet at $x=0.15$, Zhang's $SO(5)$ ideas\cite{zhang} make clear
that there is still much to be learned concerning the 
problem of the near-coexistence of an antiferromagnet and a
superconductor. What is needed is an
in-depth experimental characterization of the stripe phase.

{\em Acknowledgements.} 
Support is acknowledged by the Dutch Academy of Sciences (KNAW).

\end{multicols}


\references
\bibitem{tranquada} J. M. Tranquada, Physica B, in press (cond-mat/9709325)
and ref.'s therein. 
\bibitem{zagu} J. Zaanen and O. Gunnarsson, Phys. Rev. B {\bf 40}, 7391 (1989).
\bibitem{prelov} P. Prelov\v{s}ek and X. Zotos,
                    Phys. Rev. B {\bf 47}, 5984 (1993);
                    P. Prelov\v{s}ek and I. Sega,
                    Phys. Rev. B {\bf 49}, 15241 (1994).
\bibitem{whisc} S. R. White and D. J. Scalapino, cond-mat/9705128.

\bibitem{laughl} R. J. Laughlin, unpublished.

\bibitem{emkiv} U. L\"ow {\em et. al.}, Phys. Rev. Lett. {\bf 72}, 1918 (1994).
\bibitem{nayak} C. Nayak and
F. Wilczek, Int. J. Mod. Phys. B {\bf 10}, 2125 (1996)
\bibitem{ssh} W. P. Su, J. R. Schrieffer, and A. J. Heeger,
              Phys. Rev. B {\bf 22}, 2099 (1980).
\bibitem{prelovsek} P. Prelov\v{s}ek and X. Zotos,
                    Phys. Rev. B {\bf 47}, 5984 (1993);
                    P. Prelov\v{s}ek and I. Sega,
                    Phys. Rev. B {\bf 49}, 15241 (1994).
\bibitem{morewalls} D. Poilblanc and T. M. Rice,
                    Phys. Rev. B {\bf 39}, 9749 (1989);
                    H. J. Schulz,
                    Phys. Rev. Lett. {\bf 64}, 1445 (1990);
                    M. Kato, K. Machida, H. Nakanitshi, and M. Fujita,
                    J. Phys. Soc. Jpn. {\bf 59}, 1047 (1990);
                    J. A. Verg\'{e}s, F. Guinea and E. Louis,
                    Phys. Rev. B {\bf 46}, 3562 (1992).
\bibitem{inui} M. Inui and P. B. Littlewood,
               Phys. Rev. B {\bf 44}, 4415 (1991).
\bibitem{zaol} J. Zaanen and A. M. Oles, Ann. Physik {\bf 5}, 224 (1996).
\bibitem{nayak1}
C. Nayak and F. Wilczek, Phys. Rev. Lett. {\bf 78}, 2465 (1997).
\bibitem{tranquadani} V. Sachan {\em et. al.}, Phys. Rev. B {\bf 54},
12318 (1996). 
\bibitem{zali} J. Zaanen and P. B. Littlewood,
               Phys. Rev. B {\bf 50}, 7222 (1994).
\bibitem{schulz} e. g.,  
H. J. Schulz, in {\em Correlated Electron Systems}, ed. V. J. Emery
(World Scientific, Singapore, 1993). 
\bibitem{dagotto} E. Dagotto, Rev. Mod. Phys. {\bf 66}, 763 (1994) 
and ref.'s therein.
\bibitem{viertio} See also
H. Vierti\"o and T. M. Rice,
                  J. Phys. Cond. Matter {\bf 6}, 7091 (1994).
\bibitem{eskes}
H. Eskes {\em et. al.}, Phys. Rev. B {\bf 54}, R724 (1996);
{\em ibid.}, unpublished.
\bibitem{rice} This possibility was for the first time pointed out in:
H. Tsunetsugu, M. Troyer, and T. M. Rice, Phys. Rev. B {\bf 51}, 16456 (1994).
\bibitem{dhlee} See also O. Zachar, S. A. Kivelson and V. I. Emery, in press; Yu. A. Krotov, D.-H. Lee and A. V. Balatsky, cond-mat/9705031.
\bibitem{coen} C. N. A. van Duin and J. Zaanen, Phys. Rev. Lett. {\bf 78},
3019 (1997).
\bibitem{notesr} For instance,
why is it so that $Sr$ impurities only very weakly pin stripes, although
calculations using $1 / \epsilon r$ potentials predict impurity
potentials of order 0.5 eV? See V. I. Anisimov {\em et al.}, 
Phys. Rev. Lett. {\bf 68}, 345 (1992).
\bibitem{castell}C. Castellani, C. Di Castro and M. Grilli, cond-mat/9702112
\bibitem{zahovs}  J. Zaanen, M. Horbach and W. van Saarloos, 
Phys. Rev. B {\bf 53}, 8671 (1996).
\bibitem{nispwa} J. M. Tranquada, P. Wochner and D. J. Buttrey, 
Phys. Rev. Lett. {\bf 79}, 2133 (1997). 
\bibitem{aeppli} T. E. Mason {\em et al.}, Phys. Rev. Lett. {\bf 68}, 1414
(1992).
\bibitem{emkiv1} V. J. Emery and S. A. Kivelson, this proceedings.
\bibitem{nihit}
A caviat is that this rule is violated at higher temperatures 
(P. Wochner {\em et al.}, cond-mat/9706261). However,
it should be realized that the range of validity of the above arguments
is strictly limited to zero temperature. For instance, although there is
at least in the $x=1/3$ system a large gap in the electronic spectrum at 
low temperatures, this gap apparently closes at the charge ordering 
temperature (T. Katsufuji {\em et al.}, Phys. Rev. B {\bf 54}, R14230 (1996)),
 indicating that thermal fluctuations do play a highly 
interesting role. 
\bibitem{carmine} C. Lubritto, K. Pijnenburg, and J. Zaanen, unpublished.
\bibitem{buechner} B. B\"uchner {\em et al.}, Phys. Rev. Lett. {\bf 73},
1841 (1994).
\bibitem{qulicr} S. A. Kivelson, E. Fradkin and V. J. Emery, cond-mat/9707327.
\bibitem{zhang} S. C. Zhang, Science {\bf 257}, 4126 (1997); this proceedings.
\end{document}